\documentclass[twocolumn,tighten,times]{aastex631}

\usepackage[T1]{fontenc}

\newcommand\ergs{erg~s$^{-1}$}

\shorttitle{X-ray polarimetry of 1A~0535+262 in the supercritical state}
\shortauthors{Long et al.}

\graphicspath{{./}{fig/}}

\begin{document}

\title{X-ray Polarimetry of the accreting pulsar 1A~0535+262 in the supercritical state with PolarLight}

\author{Xiangyun Long}
\affiliation{Department of Engineering Physics, Tsinghua University, Beijing 100084, China}

\correspondingauthor{Hua Feng}
\email{hfeng@tsinghua.edu.cn}

\author[0000-0001-7584-6236]{Hua Feng}
\affiliation{Department of Astronomy, Tsinghua University, Beijing 100084, China}

\author{Hong Li}
\affiliation{Department of Astronomy, Tsinghua University, Beijing 100084, China}

\author{Ling-Da Kong}
\affiliation{Institut f{\"u}r Astronomie und Astrophysik, Kepler Center for Astro and Particle Physics, Eberhard Karls, Universit{\"a}t, Sand 1, D-72076 T{\"u}bingen, Germany}

\author{Jeremy Heyl}
\affiliation{Department of Physics and Astronomy, University of British Columbia, Vancouver, BC V6T 1Z1, Canada}

\author{Long Ji}
\affiliation{School of Physics and Astronomy, Sun Yat-Sen University, Zhuhai, 519082, China}

\author{Lian Tao}
\affiliation{Key Laboratory for Particle Astrophysics, Institute of High Energy Physics, Chinese Academy of Sciences, Beijing 100049, China}

\author{Fabio Muleri}
\affiliation{IAPS/INAF, Via Fosso del Cavaliere 100, 00133 Rome, Italy}

% authors above have direct contributions to the paper
% authors below have contributions to the instrument

\author{Qiong Wu}
\affiliation{Department of Engineering Physics, Tsinghua University, Beijing 100084, China}

\author{Jiahuan Zhu}
\affiliation{Department of Astronomy, Tsinghua University, Beijing 100084, China}

\author{Jiahui Huang}
\affiliation{Department of Engineering Physics, Tsinghua University, Beijing 100084, China}

\author{Massimo Minuti}
\affiliation{INFN-Pisa, Largo B. Pontecorvo 3, 56127 Pisa, Italy}

\author{Weichun Jiang}
\affiliation{Key Laboratory for Particle Astrophysics, Institute of High Energy Physics, Chinese Academy of Sciences, Beijing 100049, China}

\author{Saverio Citraro}
\affiliation{INFN-Pisa, Largo B. Pontecorvo 3, 56127 Pisa, Italy}

\author{Hikmat Nasimi}
\affiliation{INFN-Pisa, Largo B. Pontecorvo 3, 56127 Pisa, Italy}

\author{Jiandong Yu}
\affiliation{School of Electronic and Information Engineering, Ningbo University of Technology, Ningbo, Zhejiang 315211, China}

\author{Ge Jin}
\affiliation{North Night Vision Technology Co., Ltd., Nanjing 211106, China}

\author{Ming Zeng}
\affiliation{Department of Engineering Physics, Tsinghua University, Beijing 100084, China}

\author{Peng An}
\affiliation{School of Electronic and Information Engineering, Ningbo University of Technology, Ningbo, Zhejiang 315211, China}

\author{Luca Baldini}
\affiliation{INFN-Pisa, Largo B. Pontecorvo 3, 56127 Pisa, Italy}

\author{Ronaldo Bellazzini}
\affiliation{INFN-Pisa, Largo B. Pontecorvo 3, 56127 Pisa, Italy}

\author{Alessandro Brez}
\affiliation{INFN-Pisa, Largo B. Pontecorvo 3, 56127 Pisa, Italy}

\author{Luca Latronico}
\affiliation{INFN, Sezione di Torino, Via Pietro Giuria 1, I-10125 Torino, Italy}

\author{Carmelo Sgr\`{o}}
\affiliation{INFN-Pisa, Largo B. Pontecorvo 3, 56127 Pisa, Italy}

\author{Gloria Spandre}
\affiliation{INFN-Pisa, Largo B. Pontecorvo 3, 56127 Pisa, Italy}

\author{Michele Pinchera}
\affiliation{INFN-Pisa, Largo B. Pontecorvo 3, 56127 Pisa, Italy}

\author{Paolo Soffitta}
\affiliation{IAPS/INAF, Via Fosso del Cavaliere 100, 00133 Rome, Italy}

\author{Enrico Costa}
\affiliation{IAPS/INAF, Via Fosso del Cavaliere 100, 00133 Rome, Italy}

% 250 words 
\begin{abstract}
The X-ray pulsar 1A 0535+262 exhibited a giant outburst in 2020, offering us a unique opportunity for X-ray polarimetry of an accreting pulsar in the supercritical state. Measurement with PolarLight yielded a non-detection in 3-8 keV; the 99\% upper limit of the polarization fraction (PF) is found to be 0.34 averaged over spin phases, or 0.51 based on the rotating vector model. No useful constraint can be placed with phase resolved polarimetry. These upper limits are lower than a previous theoretical prediction of 0.6--0.8, but consistent with those found in other accreting pulsars, like Her X-1, Cen X-3, 4U~1626$-$67, and GRO~J1008$-$57, which were in the subcritical state, or at least not confidently in the supercritical state, during the polarization measurements. Our results suggest that the relatively low PF seen in accreting pulsars cannot be attributed to the source not being in the supercritical state, but could be a general feature. 
\end{abstract}

\section{Introduction}

Accreting X-ray pulsars are powered by mass accretion onto strongly magnetized neutron stars \citep[for a review see][]{Mushtukov2022}. The high magnetic pressure truncates the accretion flow at a large radius and forces the accreting materials to fall onto the star surface along magnetic field lines.  When the accretion rate or luminosity is higher than a critical value, the accretion flow will be significantly slowed down by radiation pressure above the star surface, forming radiation dominated shocks and an accretion column \citep{Basko1976}. In this supercritical state, the emission is argued to escape in form of a fan beam \citep{Becker2007}. 

\citet{Caiazzo2021} calculated the X-ray polarization signature for luminous accreting pulsars based on the accretion column model described in \citet{Becker2007}, suggesting that the geometry of the system can be inferred via phase resolved polarimetry. They predicted a polarization fraction (PF) as high as 0.6--0.8 generally. However, Observations of Her X-1 with the Imaging X-ray Polarimetry Explorer (IXPE) detected a PF of only about 0.05--0.15 over spin phases, far below the theoretical prediction \citep{Doroshenko2022}.  The authors proposed several possible scenarios to account for the discrepancy, including one that questions the presence of the accretion column, because Her X-1 displays a positive correlation between the cyclotron line energy and luminosity \citep{Staubert2014}, indicating that the source is in the subcritical state.  Cen X-3 is the second accreting pulsar observed with IXPE, which made a non-detection in its low state and measured a similar level of polarization in a bright state, with a PF of 0.05--0.15 as a function of spin phase \citep{Tsygankov2022}. It is uncertain whether the source is in the sub or supercritical state, as the cyclotron line energy does not vary as a function of luminosity \citep{Bachhar2022,Yang2023}; the X-ray luminosity during detection also gives an ambiguous determination \citep{Becker2012,Mushtukov2015}. Later, a non-detection was obtained with an IXPE measurement of 4U~1626$-$67, a persistent ultracompact low-mass X-ray pulsar, at a subcritical luminosity \citep{Marshall2022}. Recently, a similar polarization signature with PF varying around 0.1 with spin phase is seen in a transient Be/X-ray binary pulsar GRO~J1008$-$57 during its subcritical outburst in late 2022 \citep{Tsygankov2023}.

Thus, it would be interesting to measure X-ray polarization from a source in a \textit{bona fide} supercritical state, as a direct test to argue for or against this scenario.  The 2020 giant burst \citep{Kong2021,Kong2022,Ma2022,Wang2022,Chhotaray2023} from the transient Be/X-ray binary pulsar 1A~0535+262 offered us such an opportunity and was observed with the X-ray polarimeter PolarLight \citep{Feng2019}.  The outburst is the most luminous one known so far for this source and reached a peak luminosity of $1.2 \times 10^{38}$~\ergs\ \citep{Kong2021}. Observations with Insight-HXMT \citep{Zhang2020} revealed that the source transitioned into the supercritical state, where the cyclotron absorption line energy is negatively correlated with luminosity, above the critical luminosity of $6.7 \times 10^{37}$~\ergs\ \citep{Kong2021}.  The PolarLight observations cover the supercritical cycle. In this paper, we present the polarimetric results in this state and try to address the above question. 

\section{Observations and analysis}

%%%%%%%%%%%%%%%%%%%%%%%%%%%%%%%%%%%%%%%%%%%%%%%%
\begin{figure}
\centering
\includegraphics[width=\columnwidth]{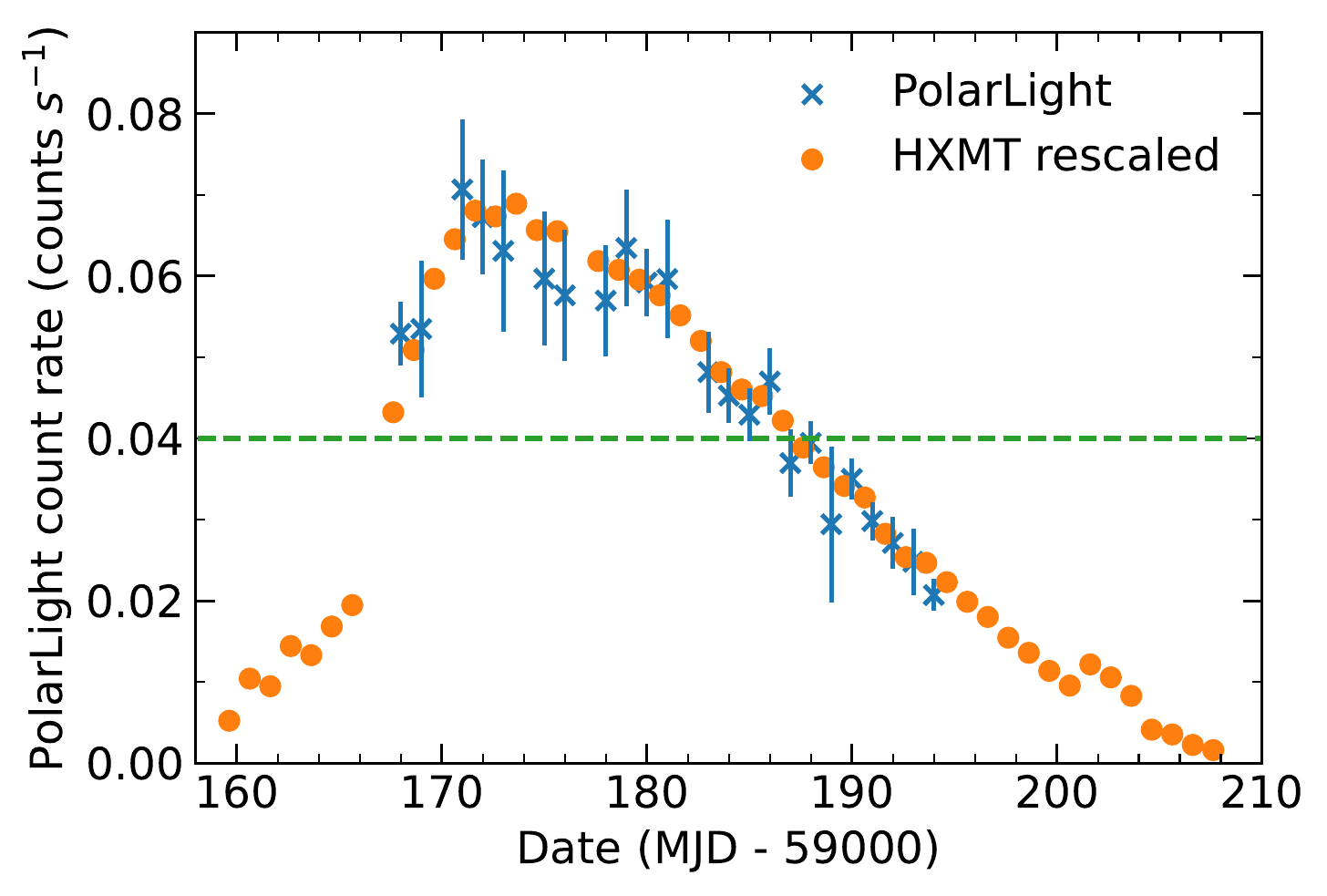}
\caption{One-day averaged 3--8 keV X-ray lightcurves of 1A~0535+262 during the 2020 major outburst measured with PolarLight (blue) and HXMT (orange; rescaled to match the PolarLight flux). The error bar has a size of 1$\sigma$ for PolarLight and is smaller than the symbol for HXMT. The horizontal dashed line marks the level above which the source was in the supercritical state \citep{Kong2021}.}
\label{fig:lc}
\end{figure}
%%%%%%%%%%%%%%%%%%%%%%%%%%%%%%%%%%%%%%%%%%%%%%%%

PolarLight is an X-ray polarimeter onboard a CubeSat \citep{Feng2019}, to measure X-ray polarization using the gas pixel detector, a 2D position sensitive proportional counter \citep{Bellazzini2013,Li2015}. It was launched in October 2018 into a Sun synchronous orbit for technical demonstration \citep{Li2021} and preliminary science observations \citep{Feng2020,Long2021,Long2022}. PolarLight observed 1A~0535+262 from November 15 to December 12, 2020, with a total effective exposure of 69~ks. The emission state and pulsar ephemeris are adopted from HXMT observations with obsIDs P0304099001-18 and P0314316001-15 \citep{Kong2021,Kong2022,Hou2023}; details about the HXMT data reduction can be found in \citet{Kong2021}. The lightcurves of the outburst measured with the two instruments are displayed in Figure~\ref{fig:lc}.  The supercritical state ranges from MJD 59167 to 59187, during which there are 33~ks of PolarLight observations in total. In this work, we focus on the analysis of this part of data. 

The energy calibration for PolarLight is implemented by comparing the observed energy spectrum with a simulated one, given the best-fit spectral model obtained with HXMT \citep{Kong2021}. With a phenomenological energy spectrum as an input, we simulate the measured spectrum with the Geant4 package and find the detector gain by comparing it with the measured spectrum. This technique has been justified in previous observations \citep{Feng2020,Li2021,Long2021,Long2022}. Events located in the central $\pm$7 mm region with an image spreading over at least 58 pixels are selected for polarimetric analysis. We employ an energy-dependent algorithm \citep{Zhu2021} to discriminate the background events. The Stokes parameters are used to calculate the polarization \citep{Kislat2015,Mikhalev2018}, and the intrinsic PF and polarization angle (PA) are inferred using a Bayesian approach to remove the bias in the estimate \citep{Maier2014,Mikhalev2018}. The peak location on the marginalized posterior distribution is adopted as the point estimate, and the credible interval (the highest density interval from the same distribution) is adopted as the error. 

We calculated the X-ray polarization in different energy bands and always got a non-detection. We thus adopted the energy band of 3--8 keV to report the result; consistent results can be obtained in other bands with larger uncertainties. In this energy band, the mean polarization modulation factor ($\mu$) is 0.42 weighted by the measured X-ray energies, and the residual background after discrimination has a fraction of 6.8\% averaged over the supercritical state. Given the number of detected photons from the source ($n_{\rm s}$) and background ($n_{\rm b}$), the minimum detectable polarization at 99\% confidence level \citep[${\rm MDP}_{99} = \frac{4.29}{\mu n_{\rm s}}\sqrt{n_{\rm s} + n_{\rm b}}$;][]{Weisskopf2010} is 0.27.

%%%%%%%%%%%%%%%%%%%%%%%%%%%%%%%%%%%%%%%%%%%%%%%%
\begin{figure}
\centering
\includegraphics[width=\columnwidth]{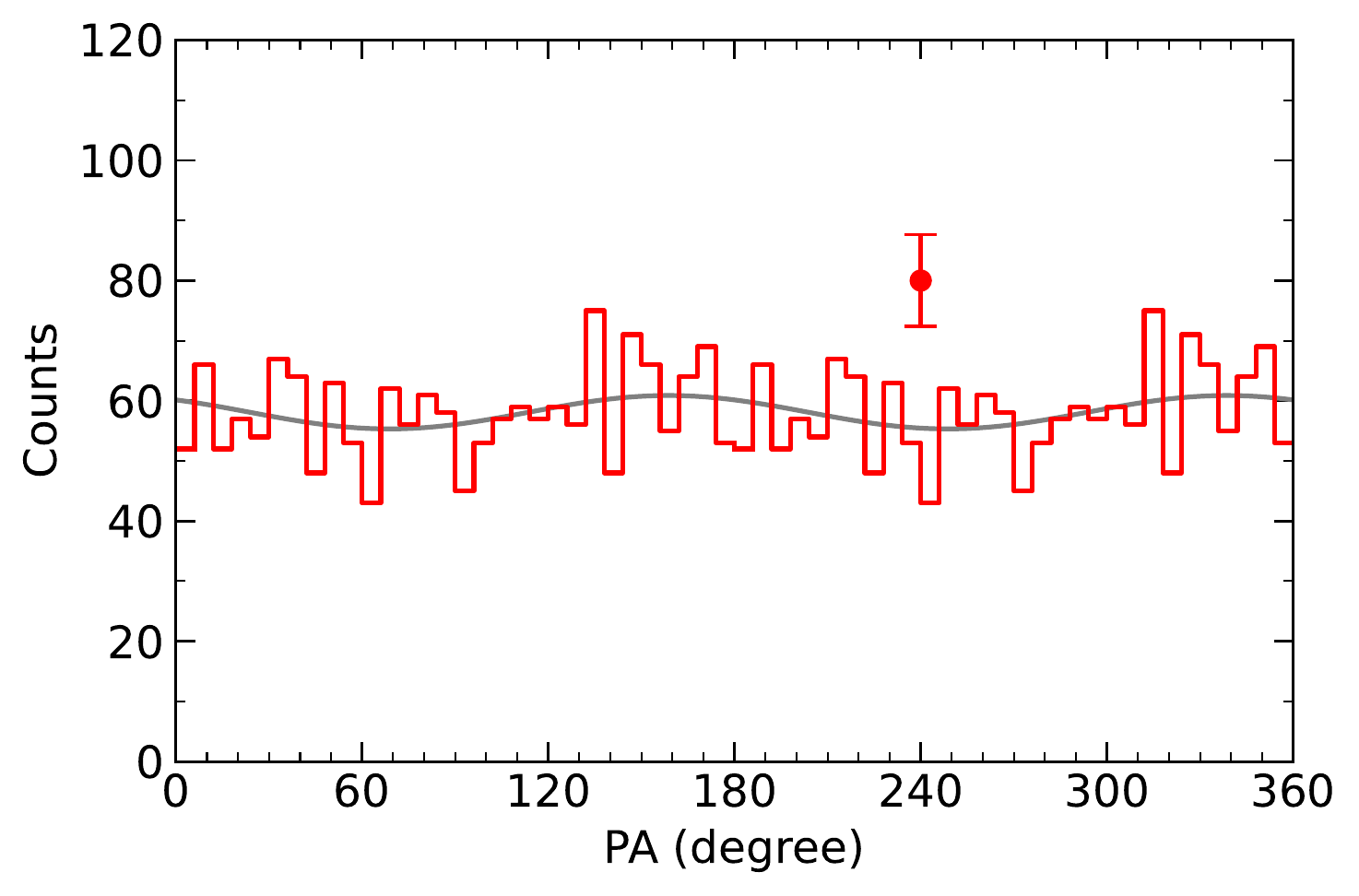}
\caption{Spin phase averaged polarimetric modulation curve of 1A~0535+262 measured with PolarLight over the supercritical state. The vertical bar indicates a typical 1$\sigma$ error. The gray line is the model curves derived from the Stokes/Bayesian analysis.}
\label{fig:mod}
\end{figure}
%%%%%%%%%%%%%%%%%%%%%%%%%%%%%%%%%%%%%%%%%%%%%%%%

%%%%%%%%%%%%%%%%%%%%%%%%%%%%%%%%%%%%%%%%%%%%%%%%
\begin{figure}
\centering
\includegraphics[width=\columnwidth]{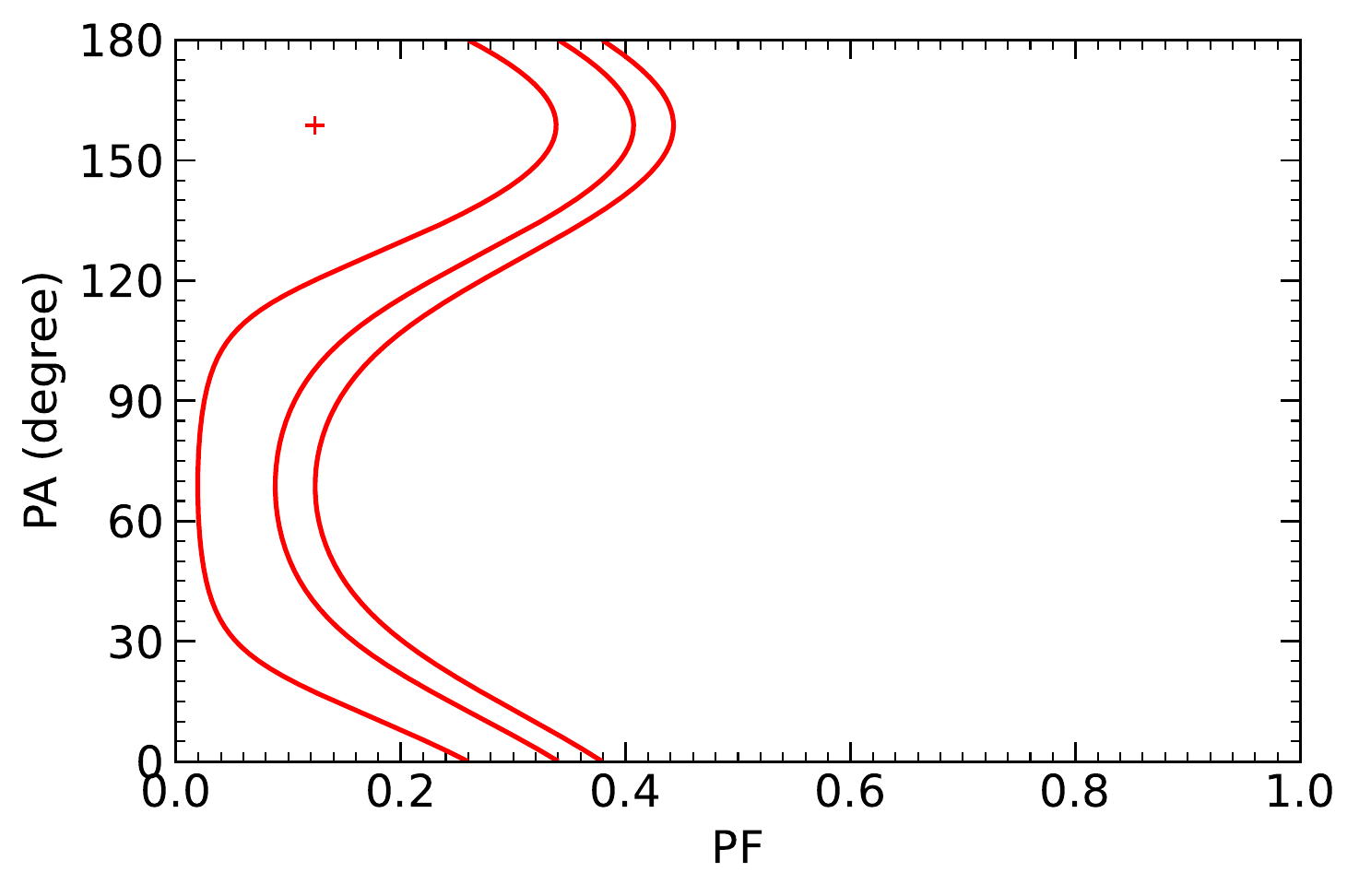}
\caption{Spin phase averaged PA vs.\ PF contours of 1A~0535+262 measured with PolarLight over the supercritical state.  The cross indicates the point estimate and the contours encircle the 90\%, 99\%, and 99.73\% credible intervals of the Bayesian posterior distribution.}
\label{fig:cont}
\end{figure}
%%%%%%%%%%%%%%%%%%%%%%%%%%%%%%%%%%%%%%%%%%%%%%%%

%%%%%%%%%%%%%%%%%%%%%%%%%%%%%%%%%%%%%%%%%%%%%%%%
\begin{figure}
\centering
\includegraphics[width=\columnwidth]{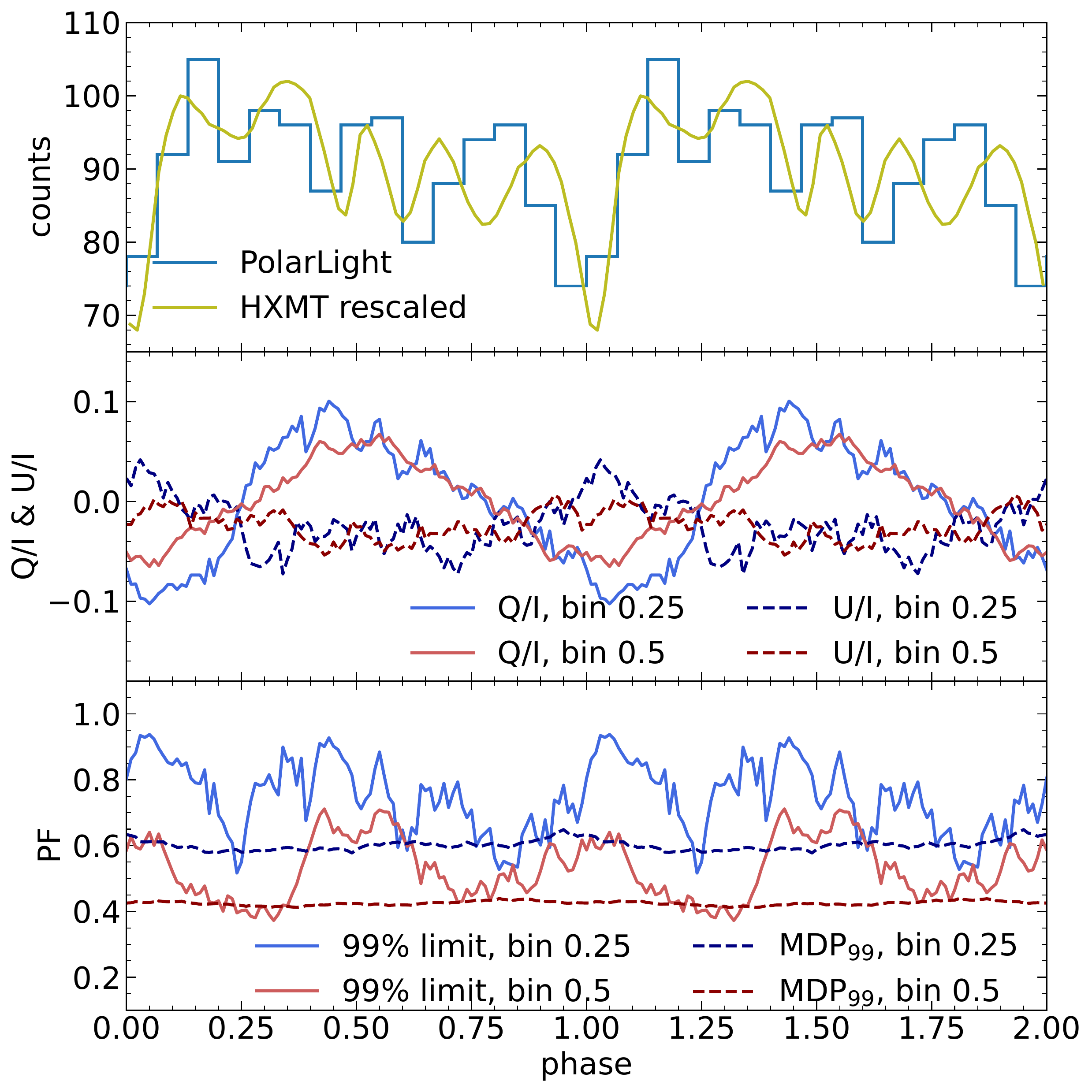}
\caption{Spin phase resolved polarimetry of 1A~0535+262 over the supercritical state in the 3--8 keV energy range. \textbf{Top}: folded pulse profiles using the ephemeris constructed with HXMT. \textbf{Middle}: normalized Stokes $Q$ and $U$ parameters as a function of phase. \textbf{Bottom}: MDP$_{99}$ and the 99\% PF upper limit as a function of phase. The polarimetric parameters are calculated by moving a phase bin with a size of 0.25 and 0.5, respectively, at a step of 0.01.}
\label{fig:phase}
\end{figure}
%%%%%%%%%%%%%%%%%%%%%%%%%%%%%%%%%%%%%%%%%%%%%%%%

In the energy band of 3--8 keV and during the supercritical state, the spin phase averaged polarization measurement gives ${\rm PF} = 0.12_{-0.12}^{+0.13}$ (90\% C.~L.), indicative of a non-detection.  We thus do not quote the measured PA. The 90\%, 99\%, and 99.73\% upper limit on the PF is 0.25, 0.34, and 0.38, respectively.  The modulation curve is displayed in Figure~\ref{fig:mod} for visual inspection.  The PA vs.\ PF posterior distribution is shown in Figure~\ref{fig:cont}.  

The PolarLight data alone do not allow us to find the pulsar spin period and fold the lightcurve. We thereby rely on the ephemeris computed with HXMT observations \citep{Kong2022,Hou2023}. Both the barycentric correction and binary motion are taken into account. The mean spin period is roughly 103.5~s and varies slowly during the outburst. Therefore, for non-simultaneous observations, we extrapolated the HXMT result from the closest epoch, typically with a time separation of 0.2--0.5~d. The pulse profiles averaged over the supercritical state measured with the two instruments are shown in Figure~\ref{fig:phase}.

We calculated the polarization at different phases. Limited by the number of photons, we chose a phase bin size of 0.25 and shift the bin center at a step of 0.01. Again, we had non-detections at any central phases; the highest detection significance is only 2.6$\sigma$.  The PF upper limits as well as MDP$_{99}$ at different central phases are also displayed in Figure ~\ref{fig:phase}. As a result of small number of photons in a phase bin, the upper limits are not constraining. 

\begin{figure}
\centering
\includegraphics[width=\columnwidth]{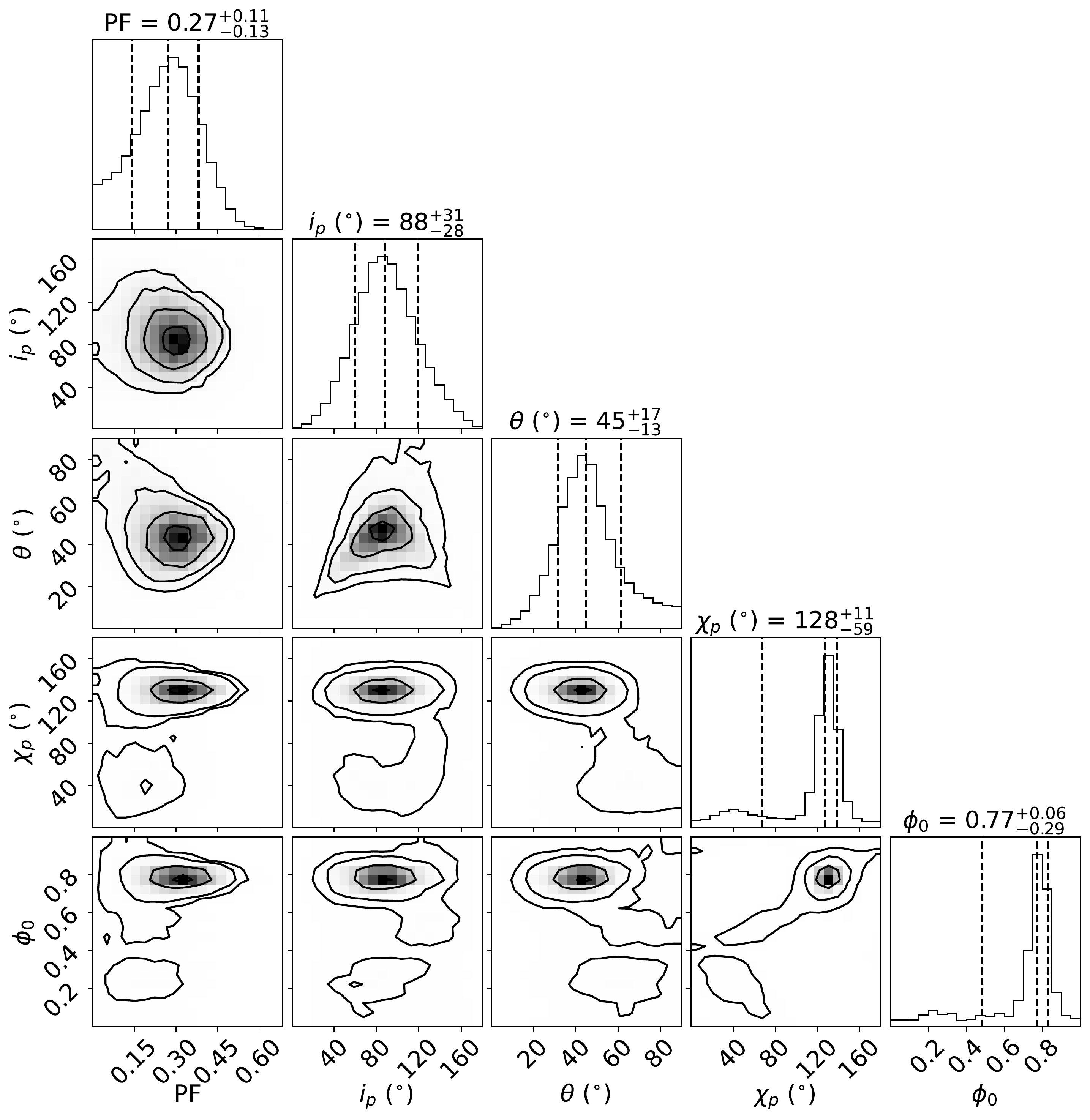}
\caption{Corner plot of the posterior distributions for the parameters in the RVM model, which assumes a constant polarization modulated by the spin of pulsar. The errors are quoted at 68\% credible level.}
\label{fig:corner}
\end{figure}

We then fit the PolarLight data with a rotating vector model \citep[RVM;][]{Radhakrishnan1969,Poutanen2020} using an unbinned likelihood analysis method \citep{Gonzalez-Caniulef2023}. This is to assume that the PA is determined by the magnetic axis of the pulsar and modulated on the sky plane by pulsar spin. Even with the presence of vacuum birefringence due to strong magnetic fields, in which case the polarization is determined by the field lines at the limiting radius far from the star surface, the RVM is still applicable if the bipolar field is the dominant component at that radius \citep{Heyl2000}. In fact, the RVM model has been successfully used to model the phase dependent PA variation in several X-ray accreting pulsars \citep{Doroshenko2022,Tsygankov2022,Tsygankov2023}. In addition to PF, the model contains the phase zeropoint $\phi_0$ and 3 parameters for the pulsar geometry: inclination of the spin axis to the line of sight $i_{\rm p}$, magnetic obliquity $\theta$, and the spin axis projected on the sky plane $\chi_{\rm p}$ \citep[see Fig.~4 in][]{Doroshenko2022}. We assume uniform priors, PF and $\phi_0$ in 0--1, $i_{\rm p}$ and $\chi_{\rm p}$ in 0--180\arcdeg, and $\theta$ in 0--90\arcdeg. With Markov Chain Monte Carlo simulations, the 90\%, 99\%, and 99.73\% upper limits on the PF are inferred to be 0.41, 0.51, and 0.57, respectively. The posterior distributions of the parameters are shown in Figure~\ref{fig:corner}. Taking into account the large uncertainties, the inferred $i_{\rm p}$ deviates from the binary inclination angle by 2$\sigma$ \citep[$88_{-28}^{+31}$\arcdeg\ vs.\ 35\arcdeg;][]{Giovannelli2007}, and the inferred $\theta$ is in line with the magnetic obliquity derived from pulse profile modeling ($45_{-13}^{+17}$\arcdeg\ vs.\ 50\arcdeg) that assumes an inclination $i_{\rm p} = 35\arcdeg$ \citep{Caballero2011,Hu2023}. If we fix $i_{\rm p}$ at 35\arcdeg, the constraint on $\theta$ is even worse. We want to emphasize that the apparent agreement is mainly because of the larger errors from our fits as a result of non-detection. Thus, we do not further discuss the geometries to avoid over-interpretation,  but focus on the PF upper limit.

%%%%%%%%%%%%%%%%%%%%%%%%%%%%%%%%%%%%%%%%%%%%%%%%
%%%%%%%%%%%%%%%%%%%%%%%%%%%%%%%%%%%%%%%%%%%%%%%%

\section{Discussion}

We present X-ray polarimetric results of the accreting pulsar 1A~0535+262 during the supercritical state. Measurements with PolarLight yielded a non-detection in 3--8 keV with a phase averaged PF upper limit of 0.34 at 99\% credible level. Due to the small area of PolarLight, the phase resolved polarimetry leads to loose constraints; we cannot rule out that the source has high PFs in some spin phase ranges.  Based on the RVM with a constant PF, we obtained a 99\% PF upper limit of 0.51.  

According to previous IXPE measurements, the RVM was always able to fit the data adequately \citep{Doroshenko2022,Marshall2022,Tsygankov2022,Tsygankov2023}. Here, based on the RVM, we are unsure if the intrinsic PF of the source is higher than those derived in other sources, but it is indeed lower than the theoretical prediction of 0.6--0.8 \citep{Caiazzo2021}, suggesting that the relatively low PFs measured in Her X-1 \citep{Doroshenko2022}, Cen X-3 \citep{Tsygankov2022}, 4U~1626$-$67 \citep{Marshall2022}, and GRO~J1008$-$57 \citep{Tsygankov2023} cannot be attributed to the sources not being in the supercritical state. One may consider other possibilities for relatively low polarizations as proposed by \citet{Doroshenko2022} and \citet{Tsygankov2022}, including mode conversion due to vacuum resonance in the emitting plasma with a strong temperature gradient \citep{Pavlov1979,Lai2002,Doroshenko2022}, the quasi-tangential effect occurred during radiation transfer \citep{Wang2009,Caiazzo2021}, and reflection/scattering on the ambient surface or medium \citep{Tsygankov2022}. To further investigate this problem, joint spectroscopy, timing, and polarimetry are needed for a large sample of accreting pulsars; this can be accomplished with the future enhanced X-ray Timing and Polarimetry (eXTP) observatory \citep{Zhang2019}.

We note that 1A~0535+262 and GRO~J1008$-$57 are transient Be/X-ray binaries \citep{Reig2011}, while Her X-1 and Cen X-3 are persistent sources; Her X-1 contains an intermediate mass companion \citep{Middleditch1976} and Cen X-3 contains a massive supergiant companion \citep{Hutchings1979}. 4U~1626$-$67 is a transient in a ultracompact binary with an extremely low mass companion \citep{Middleditch1981}. All these imply that a relatively low PF in accreting pulsars could be a general feature, independent on their companion type, accretion mode, or luminosity.

\begin{acknowledgments}
We thank the anonymous referee for useful comments. We acknowledge funding support from the National Key R\&D Project under grant 2018YFA0404502, the National Natural Science Foundation of China under grants Nos.\ 12025301, 12103027, 11821303, \& 12122306. HF acknowledges the Tsinghua University Initiative Scientific Research Program. LT acknowledges the CAS Pioneer Hundred Talent Program Y8291130K2.
\end{acknowledgments}
 
\facility{PolarLight, HXMT} 

%\bibliographystyle{aasjournal}
%\bibliography{a0535}

\end{document}